\documentclass[default,iicol]{sn-jnl}
\usepackage{amsmath, amsthm, amscd, amsfonts, amssymb, graphicx, color, subcaption, longtable,pict2e}
\usepackage{caption}
\usepackage{mathtools}
\usepackage{hyperref}
\usepackage{MnSymbol}
\usepackage{float}
\usepackage{doi}
\usepackage{listings}
\usepackage{appendix}
\usepackage{siunitx}
\usepackage{enumitem}
\usepackage{tablefootnote}
\RequirePackage{eso-pic}
\RequirePackage{graphicx}

\usepackage[absolute,overlay]{textpos} 
\usepackage{fancybox} 

\AddToShipoutPicture{%
	\AtPageUpperLeft{%
		\raisebox{-\height}{\includegraphics[height=1.2cm]{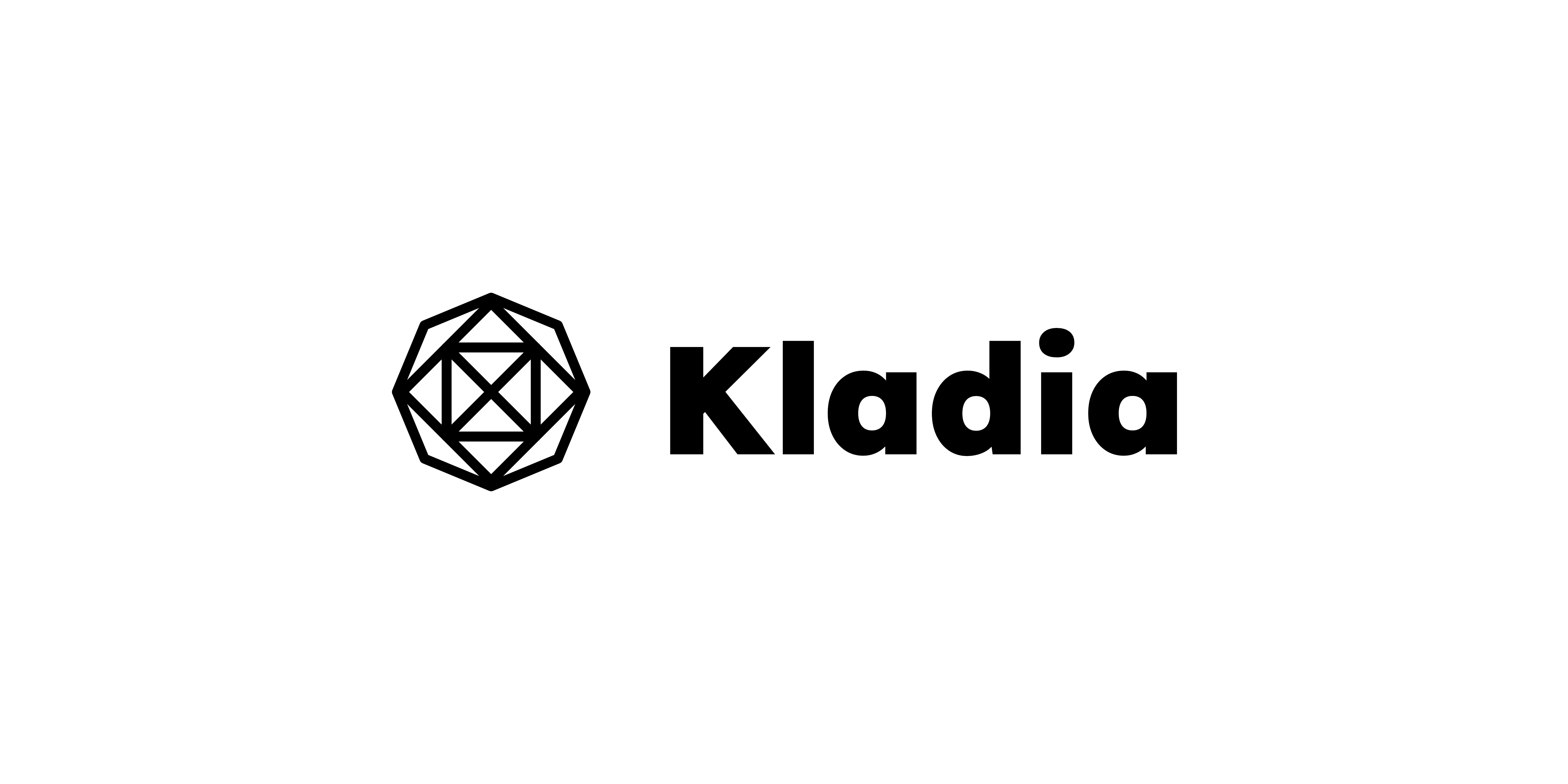}}%
	}%
}

\jyear{2021}%

\theoremstyle{thmstyleone}%
\theoremstyle{thmstyletwo}%

\theoremstyle{thmstylethree}%

\numberwithin{equation}{section}
\newtheorem{solution}{Solution}[section]%
\newcommand{\Smax}{S_{\max}}
\newcommand{\Scirc}[1]{S_{#1}}           
\newcommand{\Bgross}[1]{B_{#1}}          
\newcommand{\Bcum}[1]{B^{\mathrm{cum}}_{#1}} 
\newcommand{\Igross}[1]{I^{\mathrm{gross}}_{#1}} 
\newcommand{\Sunburn}[1]{S^{\mathrm{unburn}}_{#1}} 
\begin{document}
	\begin{textblock}{50}(20,50) 
	\fbox{\parbox{150mm}{\tiny
			\textbf{Disclaimer:}\\
			\begin{itemize}
				\item The whitepaper is for technical and community review. It is not an offer, solicitation, or investment advice.
				\item KLD does not represent a claim on any government debt. The debt index is a policy input only (indexed,
				not backed).
				\item Token price is unpredictable. Nothing in this document should be read as a promise or implication of
				profit.
				\item Some features described depend on XRPL network amendments and ecosystem availability. If a
				dependency is not active at launch, KLD will use the stated fallback process or postpone that feature.
			\end{itemize}
	}}
\end{textblock}
\title[KLD on XRPL]{Kladia Liquidity Deflator (KLD): A Debt-Indexed Deflationary Token on XRPL}


\author*[1]{\fnm{Kiarash} \sur{Firouzi}}\email{contact@kladia.com}

\author[1]{ \fnm{Parham} \spfx{} \sur{Pajouhi}}


\affil[]{\orgaddress{\city{Stockholm}, \country{Sweden}}}

	



\abstract{
	Kladia Liquidity Deflator (KLD) is an XRPL‑based, debt‑indexed token whose supply dynamics respond directly to a debt index derived from macroeconomic data sources. The model links indebtedness to deterministic adjustments in issuance, burns, and escrow release caps, creating a rule‑based deflationary mechanism that strengthens as debt rises. With a fixed maximum supply of 10\,billion KLD, the mechanism is implemented through XRPL oracles and governance. Escrow locking depends on the TokenEscrow amendment; until it is active network‑wide, allocations will be secured in a multi‑signature vault with published rules and public monitoring. KLD provides a transparent and mathematically grounded framework for a macro‑responsive digital asset.
}




\maketitle
\tableofcontents \newpage
	\section{Introduction}
	
Global debt has risen to multiples of world production in recent decades. High and rising debt levels are linked to macroeconomic fragility, currency debasement risk, as well as non-linear responses to shocks \cite{davoodi2022macroeconomic}. Sovereign, corporate, and household liabilities increase to fund fiscal programs, investments, and consumption in addition to fulfilling existing obligations \cite{borowicz2025law}. In fiat regimes, the expansion of the money supply (through credit creation and policy actions) is both a cause and an effect of rising debt \cite{dowd2012coming}. This structural feedback causes inflation to be recurrent and disinflation to be episodic, especially during stressful times. Debt necessitates liquidity, and liquidity creates more debt \cite{mitra2024stress}. These facts gradually lower the purchasing power of fiat money and erode trust in monetary authorities and intermediaries. Erosion is a common occurrence in countries with volatile currencies: saving becomes dangerous, salaries fall behind price levels, and asset allocation becomes defense \cite{jessop2013credit, d2022book}.

Cryptocurrencies have become counterpoints due to their promise of programmable monetary policy and sovereignty. However, many designs are either orthogonal to macro facts (fixed issuance schedules that never react to debt/inflation data) or vulnerable to speculative excess (hyper-inflationary emissions, unsustainable yield games, and burns detached from fundamentals) \cite{huang2022digital}. There are deflationary tokens, but their mechanisms are often arbitrary, and because their scarcity is unrelated to measurable macro indicators, it lacks an economic anchor \cite{makridis2025toward}.

\textbf{KLD} is intended to be a token whose economic policy is specifically linked to a debt metric. KLD uses debt level as a state variable in its supply rule rather than treating it as background noise. When debt level is higher relative to baseline, KLD becomes more scarce and deflationary; higher debt levels relative to the baseline correspond to tighter policy settings. The goal is to develop a token that is structurally more neutral when systemic leverage is low and structurally scarce in highly indebted regimes. 

The debt index is not meant to be a direct indicator of inflation or declining purchasing power. Inflation can rise for reasons unrelated to debt dynamics, and debt levels can rise without immediate inflationary pressure. Rather, the index should be viewed as a slow-moving indicator that represents the accumulation of financial obligations throughout the world economy, serving as a stand-in for systemic leverage and macroeconomic stress. Instead of using this index as a short-term inflation gauge or price predictor, the mechanism uses it as an open, rule-based indicator of long-term structural conditions.

XRPL was chosen to carry out the concept due to its native tokenization primitives, low-latency consensus, transaction cost efficiency, and ecosystem maturity \cite{xrpl_consensus}. Beyond narrative, the design integrates utility through a marketplace whose fees feed the deflation engine through token burns, staking to reduce float and align incentives, gamified participation to promote macro-literacy, and governance of deflation parameters and treasury policies. Our objective is to develop a token that is not only scarce but also used, and earned.

On the XRP Ledger (XRPL), KLD is implemented as an MPT issuance. The fundamental reasoning behind the model's design is as follows:
	\begin{enumerate}[noitemsep]
		\item \textbf{Mathematically Well-defined}: no negative supplies, no unbounded debt levels, and monotonic responses to debt.
		\item \textbf{Operationally Realizable}: can be implemented with XRPL primitives (token escrows, oracles, multisignature control, governance).
		\item \textbf{Economically Meaningful}: credible scarcity in high-debt regimes and transparent rules that market participants can verify.
	\end{enumerate}
	
	\section{Conceptual design}
	
	\subsection{KLD’s purpose}
	
	KLD’s primary purpose is to act as a \textbf{macro debt-indexed asset}:
	\begin{itemize}[noitemsep]
		\item It reacts to changes in a debt index derived from publicly available data (e.g. normalized debt‑to‑GDP ratios relative to baseline, total debt level).
		\item It adjusts supply-side variables (net issuance, burn, and release caps) as a deterministic function of the debt index.
		\item It offers holders exposure to a rule-based scarcity mechanism that tightens as we get higher debt levels relative to the baseline.
	\end{itemize}
	Secondary purposes of the token are:
	\begin{itemize}[noitemsep]
		\item \textbf{Utility}: Covering fees, giving access to services, and protocol usage.
		\item \textbf{Incentives}: Staking rewards and ecosystem grants.
		\item \textbf{Governance}: Parameter adjustments, oracle selection, and treasury policy.
	\end{itemize}
	
	\subsection{Economic role and sources of demand}
	
	\noindent\textbf{Demand Beyond Scarcity.}
Although it does not produce economic value on its own, the macro-indexed scarcity rule describes how KLD supply reacts to long-term structural conditions. Demand for KLD must come from practical applications that endure regardless of market sentiment if it is to remain relevant over time.
	
	\noindent\textbf{Functional Demand Channels.}
	Three categories of actual economic use are the main focus of KLD's design:
	\begin{itemize}
		\item \textbf{Protocol utility:} Participation in protocol-level activities, such as governance, staking for the validation of published macro data, and access to specific ecosystem services, requires KLD. Baseline transactional demand is produced by these uses.
		\item \textbf{Ecosystem activity:} Marketplaces, data services, and settlement tools are examples of applications built around the protocol that use KLD for fees, access, or collateral. Demand does not scale with speculative cycles, but rather with the activity of these applications.
		\item \textbf{Credibility through transparency:} A predictable monetary structure is produced by the combination of fixed supply, irreversible burns, and rule-based adjustments. For users who need stability and verifiability, this promotes long-term confidence even though it does not guarantee value.
	\end{itemize}
	
	\noindent\textbf{Limitations.}
	The mechanism does not assert that value is created solely by macro-indexed scarcity. Market adoption is contingent upon whether the ecosystem generates services that consumers deem economically beneficial or essential. Although the protocol's supply rules are intended to be clear and uniform, demand ultimately results from actual usage rather than the scarcity mechanism.

	\subsection{Debt index and external data}
\noindent\textbf{Bloc Debt Index (Annual).}
KLD references an annual Bloc Debt Index (BDI) computed from a fixed set of high-transparency blocs: 
United States, Euro Area (EA20)\footnote{For the Euro Area bloc, we use the IMF WEO ``Euro Area'' aggregate series as published in each annual cycle; any membership changes are treated as source revisions and apply prospectively only.}, Japan, United Kingdom, Canada, Australia, and South Korea (Kladia 
Core 7, KC7). For each bloc, we use “general government gross debt (\% of GDP)” from the IMF World 
Economic Outlook (WEO) dataset as the canonical source \cite{imf_weo}. The BDI is computed as a GDP-weighted average 
of bloc debt-to-GDP ratios ($D_b(t)$): 
\[
BDI(t) = \sum_b W_b(t) \cdot D_b(t),
\]
where the GDP weights are defined as:
\[
W_b(t)=\frac{GDP_b(t)}{\sum_j GDP_j(t)},
\]
using nominal GDP values from the same WEO vintage used for \(D_b(t)\). Weights are recomputed annually by this formula and cannot be manually adjusted.

The BDI is published once per year using a deterministic snapshot rule tied to the latest WEO release. Data revisions never trigger retroactive policy changes; revisions apply only to the next annual cycle. If any input is missing or disputed, the system defaults to the last confirmed value and unresolved disputes default to no execution of the proposed annual update, and the protocol continues under the last confirmed $g_t$. The bloc list, metric definition, canonical source, aggregation formula, and revision 
rule are constitutional and cannot be changed by governance.


\noindent\textbf{Revision Policy.} Macro data is subject to revisions. The index is updated only when the official source revises its published series. No retroactive adjustments are applied to past token issuance or burns; revisions affect future calculations only.\\

\noindent\textbf{Disagreement Policy.}
No governance action is taken to manually override discrepancies; the system defaults to the last confirmed value until official data is finalized.
\\

\noindent\textbf{Update Frequency.}
The KC7 Bloc Debt Index is updated once per year using the latest available IMF World Economic Outlook (WEO) database designated for the annual cycle. The annual snapshot is published according to the deterministic snapshot rule in Table \ref{t2}. Within a given year, the policy factor $g_t$, which we define in section \ref{sec3}, remains constant; monthly escrow release caps and annual issuance budgets reference the same annual $g_t$.
\\

\noindent\textbf{Governance Scope.} The metric, the sources, and the frequency of updates cannot be altered by governance. The KC7 bloc list, metric definition, canonical source, aggregation formula, annual cadence, and revision rule are constitutional and cannot be changed by governance. If a future index design is desired, it will be introduced only through a new token version, not by modifying KC7 rules.
\\

   From these, our protocol defines a \textbf{debt index} \(X_t\). Conceptually \(X_t\) represents the \emph{normalized debt‑to‑GDP ratios relative to baseline} for the KC7 blocs. The mechanism does not react to short-term fluctuations, but rather to the \emph{level} of the debt ratio in relation to a fixed baseline year. The policy tightens when the debt ratio is higher than the baseline and relaxes when it is lower. This rule is not growth-based; rather, it is level-based.
	
	\section{Mathematical model}
	The debt-indexed framework that controls KLD's supply dynamics is formalized in this section. We define the debt index, present the fundamental variables, and outline the policy functions that control issuance, burns, escrow releases, and staking emissions. The model gives the token's macro-responsive behavior a clear, deterministic basis.
	
	\subsection{Debt normalization}\label{sec3}
	The normalized index $X_t$ is computed relative to the fixed KC7 Bloc Debt Index baseline. Let 
	$BDI(t)$ denote the GDP-weighted average of bloc debt-to-GDP ratios from the IMF WEO dataset, and let 
	$BDI_{\mathrm{ref}}$ denote the baseline snapshot. Then the \emph{Debt Index} is defined as:
	\[
	X_t = \frac{BDI(t)}{BDI_{\mathrm{ref}}}.
	\]
	This ensures that policy variables are scaled consistently against a constitutional, immutable bloc baseline rather than a single-country measure.
	
	Define the excess debt ratio as: 
	$$x_t=\max(0, X_t - 1).$$
	
\noindent\textbf{Baseline Period.}
The baseline reference for normalization is fixed at genesis (the timestamp of the initial on-ledger MPT issuance transaction). Let $WEO_{\mathrm{vintage},0}$ be the most recent IMF World Economic 
Outlook (WEO) dataset publicly released prior to genesis. The baseline value $BDI_{\mathrm{ref}}$ is computed from $WEO_{\mathrm{vintage},0}$ using the KC7 definition in Table \ref{t2} and is immutable thereafter. For transparency and reproducibility, the project will publish \emph{(i)} the WEO vintage identifier (release month/year), \emph{(ii)} the exact series names used, and \emph{(iii)} a cryptographic hash of the raw input file(s) referenced for $BDI_{\mathrm{ref}}$. 

Annual updates compute $BDI(t)$ using the latest WEO vintage available under the same deterministic rule. WEO revisions never rewrite historical outputs; revisions apply starting from the next annual 
cycle.

\begin{table}[h!]
	\centering
		\scalebox{0.75}{%
	\begin{tabular}{ll}
		\hline
		Parameter & Value \\
		\hline
		Bloc set (KC7) & US, EA20, JP, UK, CA, AU, KR \\
		Metric & General government gross debt (\% of GDP) \\
		Canonical source & IMF World Economic Outlook (WEO) dataset \cite{imf_weo} \\
		Aggregation formula & GDP-weighted average of bloc debt ratios \\
		Snapshot cadence & Annual (rule-based)\tablefootnote{Snapshot cadence is deterministic: Publication occurs at 12:00 UTC on the 10th calendar day after the IMF publishes the October WEO dataset. If the October dataset is not published by December 1 of that year, the protocol uses the latest publicly released WEO vintage available as of December 1, and publication occurs at 12:00 UTC on December 10 (or next business day). The Policy Report records the WEO vintage identifier, publication timestamp, and dataset hash.		
		}
		 \\
		Revision rule & No retroactive changes; revisions apply to next cycle \\
		Missing data rule & Last confirmed value, disclosed in Policy Report \\
		\hline
	\end{tabular}}
	\caption{Bloc Debt Index (Annual Snapshot)}\label{t2}
	\footnotesize{Note: KC7 bloc list, metric definition, canonical source, aggregation formula, and revision rule are constitutional and immutable.}
\end{table}

If the annual update is missing or disputed, no new annual update is executed and the protocol continues operating under the last confirmed \(g_t\) until a valid uncontested update is published (or until the next annual cycle).

	To compress \(x_t\) into \([0,1)\) and ensure bounded responses, define:
	\[
	g_t = \frac{x_t}{1 + \lambda x_t},
	\]
	where \(\lambda>0\) is a scaling parameter controlling sensitivity. The following properties hold for $g_t$:
	\begin{itemize}[noitemsep]
		\item \(g_t=0\) when \(BDI(t)\le BDI_{\mathrm{ref}}\).
		\item \(g_t \in (0,1)\) for \(x_t>0\), with \(g_t \to 1\) as \(x_t\to\infty\).
		\item \(g_t\) increases monotonically as debt rises above baseline (and hence in \(x_t\)).
	\end{itemize}
	The parameter \(\lambda\) controls how quickly \(g_t\) approaches 1 as debt rises. The debt ratio's \emph{level} in relation to a baseline year is used by the mechanism. It doesn't make use of the debt growth or change rate. This level-based structure is reflected in the mathematical definition of $g_t$.\\

		\noindent\textbf{Timing Convention.}
The policy factor $g_t$ is updated annually from the KC7 Bloc. Issuance budgets are set annually, escrow releases are capped monthly, and fee burns/staking emissions accrue continuously; all are reported in the annual Policy Report and may be summarized more frequently for transparency.

	\subsection{Supply variables and dynamics}
	Let:\\
	\begin{description}[noitemsep]
		\item[] \(S_{\max}\): Fixed maximum KLD supply:\\ \(10{,}000{,}000{,}000\) tokens.
		\item[] \(S_t\): Total released KLD at time \(t\), with \(0 \le S_t \le S_{\max}\).
		\item[] \(C_t\): Circulating supply at time \(t\) (excludes unvested and escrow-locked amounts).
	\end{description}
	
	We define the net supply change over a period as:
	\[
	\Delta S_t = I_t - B_t,
	\]
	where \(I_t\) is gross issuance (e.g., vesting, escrow releases, staking rewards), and \(B_t\) is total burns in each period \(t\). The core idea is to make both \(I_t\) and \(B_t\) functions of \(g_t\).\\
	
The maximum supply of KLD is set at $10{,}000{,}000{,}000$ tokens, all of which are generated at genesis. There will never be any more tokens created. Only the \emph{release of already-created tokens} from reserves or locked allocations into circulating supply is referred to as ``issuance," ``annual issuance budgets," or ``gross issuance." These releases only alter the distribution between circulating and locked balances; they do not raise the overall supply.
	
The MPT supply cap on XRPL restricts the maximum number of tokens that can be in circulation at any given time, but it does not ensure permanent scarcity on its own because burned tokens could theoretically be reissued up to the cap. KLD mitigates this by adopting a strict no-reissuance commitment enforced through issuer-control minimization, multisignature constraints, public monitoring, and independent audit (see Section \ref{4.5}). We implement the following rule in order to remove this ambiguity.

 KLD does not mint additional tokens after genesis. Burns are intended to be permanent and to reduce circulating supply over time. To prevent discretionary re-issuance, KLD adopts a strict no re-issuance commitment enforced through issuer-control minimization, multisignature constraints, public monitoring, and independent audit (see Section \ref{4.5}). Any post-genesis re-issuance attempt would be publicly visible on-ledger and would constitute a breach of the project’s published commitments. Any proposal to change this rule requires a new token version with a new genesis specification and independent audit.

\subsection{Policy bands for debt}
	
Governance can establish informal bands for $g_t$ for interpretability. We define three main regimes to analyze debt data:
	\begin{itemize}[noitemsep]
		\item \textbf{Low-debt regime}: \(g_t=0\) (debt near or below reference level).
		\item \textbf{Moderate-debt regime}: intermediate \(g_t\).
		\item \textbf{High-debt regime}: \(g_t\) close to 1 (debt far above reference level).
	\end{itemize}
	To prevent cliff effects, the policy itself is continuous in \(g_t\) instead of piecewise continuity.
	
	\subsection{Debt-indexed issuance policy}
	
	Let \(I^{\text{base}}\) denote a baseline annual issuance budget
	(e.g., for staking rewards and ecosystem releases).
	Define the \emph{issuance factor} as:
	\[
	\phi_I(g_t) = 1 - \alpha_I g_t,
	\]
	with \(\alpha_I \in [0,1]\).
	Define annual gross issuance as:
	\[
	I_t^{\text{gross}} = \max\left(0,\, \phi_I(g_t)\,I^{\text{base}}\right),
	\]
	subject to the cap:
\[
I^{gross}_t \le S^{locked,unburn}_t
\]
where \(S^{locked,unburn}_t\) is the remaining unburned balance held in escrow/vault allocations. Interpretation:
	\begin{itemize}[noitemsep]
		\item Lower debt levels relative to the baseline (\(g_t \approx 0\)) correspond to issuance near baseline.
		\item Higher debt levels relative to the baseline (\(g_t\to 1\)) means that issuance declines linearly, and can reach zero if \(\alpha_I=1\).
	\end{itemize}	
	Cumulative burns and remaining unburned reserves are defined as:
	\[
	\Bcum{t} := \sum_{s=0}^{t} \Bgross{s}, 
	\qquad 
	\Sunburn{t} := \Smax - \Bcum{t},
	\]
where $B^{cum}_t$ 
is the cumulative number of tokens that have been permanently burned up to and including time $t$, which aggregates all on-ledger burn events since genesis, and only increases (never decreases).	We define circulating supply recursion and issuance/burn semantics as:
	\[
	\Scirc{t+1} = \Scirc{t} + \Igross{t} - \Bgross{t}.
	\]
Therefore:
	\[
	\Scirc{t} + \Igross{t} \le \Smax 
	\quad \text{(bookkeeping upper bound)}
	\]
	\noindent\textbf{Operational constraint:} annual releases are additionally bounded by the remaining unburned tokens held in escrow/vault; releases cannot exceed locked reserves available for distribution.
	\[
	\Igross{t} \le \Sunburn{t}.
	\]
	\subsection{Debt-indexed burn policy}
	
	Let \(b^{\text{base}}\) be a baseline burn fraction of protocol fees in KLD.
	Define:
	\[
	b(g_t) = \min\left(b_{\max},\,b^{\text{base}} + \beta_B g_t\right),
	\]
	with \(\beta_B \ge 0\) and upper bound \(b_{\max} \le 1\).
	If protocol fees in period \(t\) equal \(F_t\) (in KLD), the token burn amount for period $t$ is therefore:
	\[
	B_t^{\text{fees}} = b(g_t)\,F_t.
	\]
	Higher debt level implies a higher fraction of fees is burned, increasing deflationary pressure.
	
	\subsection{Debt-indexed escrow release caps}
	
Let $E_{base}$ be the initial monthly escrow release cap. Define:
	\[
	E_{\text{cap}}(g_t) = \max\left(E_{\min},\, E_{base}\,(1 - \alpha_E g_t)\right),
	\]
	with \(\alpha_E\in[0,1]\) and floor \(E_{\min} \ge 0\).
	This cap restricts the amount of the escrowed allocation that can be distributed each month. As debt rises, the cap becomes more stringent.
	
	\subsection{Staking emission adjustment}
	
We establish a staking emission factor if a percentage of issuance is devoted to staking rewards as follows:
	\[
	\phi_r(g_t) = 1 - \gamma g_t,\qquad \gamma\ge 0,
	\]
	where $\gamma$ is the sensitivity coefficient that determines how strongly the staking emission rate reacts to debt. 

	Let \(r^{\text{base}}\) denote a baseline staking emission rate per period (fraction of staking reserve).
	The effective staking emission rate becomes:
	\[
	r(g_t) = \max(0,\,\phi_r(g_t)\,r^{\text{base}}).
	\]
	This reduces inflationary pressure from staking when debt is at higher levels, preserving scarcity. The parameter $\gamma \ge 0$ controls how strongly the staking emission rate responds to the debt index $g_t$. A larger $\gamma$ causes $r(g_t)$ to decrease more rapidly as debt is at higher levels, increasing deflationary pressure, while $\gamma=0$ makes staking emissions independent of debt.
	
	\subsection{Summary of net deflationary behavior}
	
	Combining all the results in the previous sections we get:
	\begin{itemize}[noitemsep]
		\item \(I_t^{\text{gross}}\) decreases with \(g_t\),
		\item \(b(g_t)\) increases with \(g_t\),
		\item \(E_{\text{cap}}(g_t)\) decreases with \(g_t\),
		\item \(r(g_t)\) decreases with \(g_t\).
	\end{itemize}
	As debt is at higher levels relative to baseline, all major levers shift in a way that \emph{reduces net supply growth}. In terms of debt, KLD's scarcity is therefore \emph{anti-cyclical}.\\
	
The market price of KLD is intrinsically unpredictable, even though shifts in the circulating supply can affect scarcity. Numerous external factors, such as real adoption, liquidity conditions, market structure, and more general macroeconomic dynamics, influence price formation. The mechanism outlined in this paper is not intended to target or imply any particular price outcome, but rather to modify supply in accordance with transparent rules. Regarding anticipated appreciation or future price behavior, no claims are made.

	\section{Token specification and distribution}
	
	\subsection{XRPL token specification}

	\begin{description}[noitemsep]
		\item[\textbf{Token name}:] Kladia Liquidity Deflator (KLD)\\
		\item[\textbf{Ledger}:] XRP Ledger (XRPL)\\
		\item[\textbf{Token type}:] XRPL Multi‑Purpose Token (MPT) issuance \cite{xrpl_mpt_docs})\\
		\item[\textbf{Total token supply}:] \(S_{\max}=10{,}000{,}000{,}000\)\\ KLD (immutable cap)\\
		\item[\textbf{Issuer account}:] A designated multi-signature XRPL issuer account is responsible for executing token-level actions—such as releases, burns, and lock updates—strictly according to publicly documented rules, with all operations verifiable on-chain.\\
	\end{description}

Policy logic is not carried out independently by the XRPL. All supply modifications, including burns, releases, and lock updates, are carried out by specific multi-signature accounts that strictly adhere to rules that are made public. Anyone can confirm that the multi-signature operators adhere to the established process because the debt index, policy inputs, and subsequent actions are published on-chain. Without requiring automated ledger-level execution, this structure  is designed to make actions publicly verifiable and to enforce  accountability and transparency.

\noindent\textbf{Liquidity and Market Structure.}
Initial liquidity for KLD will be provided through centralized exchanges, OTC partners, or other 
permissioned venues. Although MPTs can be marked as ``Can Trade,'' on-ledger MPT trading on the 
XRPL DEX is not currently available. Accordingly, we do not assume XRPL DEX liquidity at launch, 
and all early liquidity provisioning and price discovery will occur off-ledger until native MPT 
DEX support becomes available.

	\subsection{Initial allocation}
The initial distribution of the fixed KLD supply among key ecosystem functions is summarized in Table \ref{t1}. In order to have regulated, rule-based releases under the debt-indexed policy, the majority is locked in ledger escrows. While operational, community, staking, and liquidity reserves supply the resources required for ecosystem growth and stability, the co-founders and team allocation gradually vests to align long-term incentives.
	\begin{table}[h!]
		\centering
		\caption{KLD token allocation}\label{t1}
		\scalebox{0.75}{%
		\begin{tabular}{lrr}
			\toprule
			\textbf{Allocation} & \textbf{\%} & \textbf{Amount (KLD)} \\
			\midrule
			Escrow (on-ledger, monthly releases) & 55\% & 5,500,000,000 \\
			Co-founders, Team \& Advisors (vested) & 25\% & 2,500,000,000 \\
			Company Reserve (operational) & 10\% & 1,000,000,000 \\
			Community \& Airdrops & 5\% & 500,000,000 \\
			Staking Rewards Reserve & 3\% & 300,000,000 \\
			Liquidity \& Partnerships & 1.5\% & 150,000,000 \\
			Legal \& Treasury & 0.5\% & 50,000,000 \\
			\midrule
			\textbf{Total} & \textbf{100\%} & \textbf{10,000,000,000} \\
			\bottomrule
		\end{tabular}
	}
	\end{table}

	\subsection{Co-founders and team vesting}
	
	The co-founders, team, and advisors allocation\\ \(T=2{,}500{,}000{,}000\) KLD is subject to:
	\begin{itemize}
		\item[1.] A 12-month cliff during which no tokens are transferable.
		\item[2.] Linear monthly vesting over the following 36 months.
	\end{itemize}
	Monthly vesting amount is:
	\[
	\text{MonthlyRelease}_{\text{team}} = \frac{T}{36}
	\approx 69{,}444{,}444.44\ \text{KLD}.
	\]
	\subsection{Escrow structure}
	
	The escrowed portion, \(E_{total}=5{,}500{,}000{,}000\) KLD, will move from the temporary multi-signature vault to native token escrows once the TokenEscrow amendment is operational throughout the XRPL network \cite{xrpl_known_amendments}. Then we will have:
	\begin{itemize}[noitemsep]
		\item Monthly release constrained by \(E_{\text{cap}}(g_t)\).
		\item Ability to relock unused tokens into new escrows with hashes and justification for transactions.
	\end{itemize}
	This change enforces transparent auditability, deterministic release schedules, and direct enforcement of the escrowed supply at the ledger level. This provides on-ledger enforcement of the escrow release schedule and release caps. Governance and policy execution remain subject to the protocol’s published process, multisignature controls, transparency reporting, and the challenge-window verification system. 
	
	\subsection{Token Governance and Treasury Transparency}\label{4.5}
	
	\noindent\textbf{Purpose of the Escrow Allocation (55\%).}
	The 55\% allocation represents the long-term ecosystem and public-incentive reserve. Its sole permitted uses are:
	\begin{itemize}
		\item community incentives and grants,
		\item liquidity provisioning for exchanges and market infrastructure,
		\item ecosystem development programs,
		\item public-good funding aligned with the protocol mission.
	\end{itemize}
	No portion of this allocation may be transferred to founders, team members, or private investors.
	
\noindent\textbf{Control Structure.}
All treasury and escrow allocations are held in publicly visible XRPL accounts controlled by a multi-signature scheme. The configuration is:
\begin{itemize}
	\item 5-of-8 multisignature for the ecosystem escrow (55\%),
	\item 6-of-7 multisignature for the company reserve (10\%),
	\item 3-of-5 multisignature for team/advisor vesting wallets (25\%).
\end{itemize}
No single individual or entity can unilaterally move tokens. Minting is permanently disabled after genesis; only pre-created allocations may be released under the published caps and multisignature controls.

\noindent\textbf{Normal Release Budget.}
The ecosystem escrow has a base annual release budget of up to 5\% of the escrow balance, which is then adjusted downward by the debt pressure factor ($g_t$) and further constrained by the monthly cap ($E_{\mathrm{cap}}(g_t)$). In higher debt pressure regimes, releases decrease mechanically; in low 
debt pressure regimes, releases may approach the base budget but never exceed the hard caps.

	\noindent\textbf{Relocking of Unused Tokens.}
	``Relocking" means returning unreleased tokens back into the same escrow or multi-signature vault under the original constraints. Relocking cannot modify the vesting horizon, increase the release limit, or transfer tokens to any other wallet. It is strictly a mechanism to prevent accidental or premature circulation.\\
	
	\noindent\textbf{Relocking Constraint.}
	Relocking may only reduce near-term circulating supply and cannot be used to increase future release 
	rights beyond the published base budgets and caps. All relocks must reference the original allocation 
	category, be executed on-ledger, and be disclosed in the next Policy Report with transaction hashes 
	and justification.\\
	
	\noindent\textbf{Team and Advisor Vesting.}
	Team and advisor tokens (25\%) are held in time-locked multi-signature wallets. During vesting:
	\begin{itemize}
		\item tokens cannot be transferred or used,
		\item founders and team members have \emph{no voting power} associated with unvested tokens,
		\item governance participation requires tokens that are fully vested and circulating.
	\end{itemize}
	
	\noindent\textbf{Company Reserve Safeguards (10\%).}
	The company reserve is restricted to operational expenses. To prevent misuse:
	\begin{itemize}
		\item all transfers from the reserve require a 6-of-7 multisignature approval,
		\item the project commits to a self-imposed monthly spending guideline (e.g., not exceeding 1\% 
		of the reserve under normal conditions),
		\item all expenditures must be disclosed in the quarterly treasury report.
	\end{itemize}
	On-chain, the only enforceable control is the 6-of-7 multisignature requirement; all additional 
	policies are publicly documented commitments, visible through transparency reports.

	\noindent\textbf{Wallet Transparency.}
	The following XRPL accounts will be published prior to launch:
	\begin{itemize}
		\item ecosystem escrow wallet,
		\item company reserve wallet,
		\item team/advisor vesting wallets,
		\item operational multisig wallet for protocol actions.
	\end{itemize}
	Each wallet is permanently tagged and publicly monitored.
	
	\noindent\textbf{Reporting Cadence.}
	We commit to a recurring transparency schedule:
	\begin{itemize}
		\item \textbf{Monthly:} summary of treasury movements and operational spending,
		\item \textbf{Quarterly:} full treasury report including balances, releases, burns, and governance actions,
		\item \textbf{Annually:} independent third-party review of treasury and policy execution.
	\end{itemize}
Quarterly reports are treasury/accounting summaries only and do not trigger any mid-year update of \(g_t\).

\noindent\textbf{On-chain publication.} For each report, the protocol publishes an on-ledger commitment consisting of (i) a cryptographic hash of the full report content and (ii) a reference link (e.g., website/IPFS). The full report is archived off-chain; the on-ledger hash enables public verification that the document was not altered after publication.

	\noindent\textbf{Burn Finality and Issuance Disablement.}
Burns are intended to be permanent and to reduce the circulating supply over time. To prevent discretionary re-issuance, KLD is created at genesis and the project adopts a strict no minting rule enforced by issuer-control minimization, multisignature constraints, public monitoring, and independent audit. 
	
Immediately after genesis issuance, we will cryptographically and operationally remove the ability to perform further minting actions from the issuer configuration. The exact on-ledger configuration and verification steps will be published in the technical appendix and validated by an independent audit prior to launch. Public verification of supply integrity will remain possible through XRPL explorers and audit reports.

	\section{Oracle and data architecture}
	
	\subsection{Debt data ingestion}
	
	External data on debt is ingested by a set of oracles:
	\begin{itemize}[noitemsep]
		\item Oracles fetch periodic debt statistics from external sources.
		\item Raw values \(BDI(t)\) are normalized on-chain or off-chain into \(X_t\), then into \(g_t\).
		\item Oracle reports are signed and posted to XRPL by a cryptographic hash + a pointer (URL/IPFS) in transaction memos, forming the basis for policy updates.
	\end{itemize}
	
	\subsection{Oracle robustness}
	
	To reduce manipulation risk:
	\begin{itemize}[noitemsep]
		\item Multiple independent oracle operators provide inputs.
		\item The protocol aggregates inputs (e.g. via median) to produce a single \(BDI(t)\).
		\item Governance can replace, add, or remove oracle operators.
		\item Misbehaving oracles face reputational or contractual penalties off-chain.
	\end{itemize}
	
	\subsection{Policy Update Process}
	
	\noindent\textbf{1. Oracle Publication.}
After obtaining the official IMF WEO dataset, authorized oracle operators publish a signed on-ledger update containing:
\begin{itemize}
	\item [(i)] each bloc’s debt-to-GDP ratio $D_b(t)$, 
\item[(ii)] each bloc’s nominal GDP, $GDP_b(t)$, 
\item[(iii)] the computed Bloc Debt Index $BDI(t)$, 
\item[(iv)] the normalized index $X_t$ relative to the fixed baseline $BDI_{\mathrm{ref}}$, and 
\item[(v)] the derived policy variable $g_t$.
\item[(vi)] the WEO vintage identifier and a cryptographic hash of the raw WEO dataset file(s) used in this cycle.
\end{itemize}
	Every submission is timestamped, signed, and publicly 
	reproducible. All raw bloc inputs and computed weights are disclosed in the annual Policy Report.
	
	\noindent\textbf{2. Challenge Window.}
	After publication, a fixed 72-hour challenge window begins.
	\begin{itemize}
		\item During this period governance cannot manually override oracle values; the system defaults to no action until the window closes.
		\item the median of submitted values is displayed but not yet actionable,
		\item any oracle operator or governance participant may flag discrepancies or data issues,
		\item no policy action may be executed.
	\end{itemize}
	Individual flags are recorded as comments and do not halt execution unless they meet the valid-dispute threshold in Step 3.

	\noindent\textbf{3. Dispute Handling.}
The default result is no execution of the proposed update if a valid dispute is raised during the challenge window. The value of $X_t$ cannot be changed by governance, although it may ask for clarification or replace defective oracle operators. Until the problem is fixed and a fresh, uncontested update is released, the most recent confirmed value is still in force.
	
	A dispute is considered valid only if \emph{(a)} at least 2 independent oracle operators flag the same issue, or \emph{(b)} a governance vote meeting quorum passes a ``pause execution” motion during the challenge window. Individual flags that do not meet this threshold are recorded as comments but do not halt execution.
	
	If execution is paused, a corrected oracle update must be published within 14 days; otherwise the system continues under the last confirmed ($g_t$) until the next annual cycle.
	
	\noindent\textbf{4. Execution.}
    The policy executor multisignature wallet (5-of-8) executes the permitted actions (issuance, burns, or escrow releases) based on the published $g_t$ if the challenge window expires without dispute. Every action is carried out on-ledger and added to the upcoming transparency report.
	
	\noindent\textbf{5. Emergency Procedure.}
	Governance may halt policy execution in the event of a blatant Oracle malfunction or data outage, but they are unable to manually change or override the debt index's value. When valid Oracle data is restored, the system returns to its regular functioning.
	
	\subsection{Policy Report, Verification, and Failure Modes}
	Each policy cycle produces a signed report published using an on-ledger hash and archived off-chain (see Section \ref{4.5}). The report contains:
	\begin{itemize}
		\item the Debt Index $X_t$ and derived policy variable $g_t$,
		\item the oracle submissions and median aggregation,
		\item the governance vote outcome (if applicable),
		\item the executed actions (issuance, burns, escrow releases),
		\item the signer set and transaction hashes,
		\item the raw bloc debt and GDP inputs used for the cycle,
		\item the computed $BDI(t)$, the baseline $BDI_{\mathrm{ref}}$, and the normalized index $X_t$,
		\item the GDP weights $W_b(t)$ applied in the aggregation.
		\item the WEO vintage identifier and dataset hash used to compute \(BDI(t)\), plus any fallback note if canonical retrieval failed.
	\end{itemize}
	
	\noindent\textbf{Verification.}
	All reports are verifiable by any participant:
	\begin{itemize}
		\item oracle values and execution transactions are visible on XRPL,
		\item signatures from the multisig executor are auditable,
		\item supply changes can be reconciled against the published report.
	\end{itemize}
	
	\noindent\textbf{Challenge Window.}
	After oracle publication, a fixed challenge window (72 hours) applies before execution. 
	During this period:
	\begin{itemize}
		\item discrepancies may be flagged,
		\item disputes default to \emph{no action},
		\item governance may replace faulty oracle operators but cannot override values.
	\end{itemize}
	
	\noindent\textbf{Failure Modes.}
	If oracle malfunction or operational error occurs:
	\begin{itemize}
		\item governance may pause execution but cannot set manual values,
		\item the last confirmed index remains in effect until a valid update is published.
	\end{itemize}
	``No action” refers to no execution of the proposed annual update; it does not imply freezing all protocol activity. Releases/burns continue under the last confirmed policy setting unless execution is formally paused by quorum governance.
	
	\section{Governance}	
	This section describes the framework for decision-making that specifies how parameters are updated, how roles are assigned, and how the protocol maintains accountability and transparency over time.\\
\noindent\textbf{Immutable Parameters.}
The following elements are permanently fixed and cannot be modified by governance:
\begin{itemize}
	\item the definition of the Debt Index $X_t$,
	\item the baseline ratio $BDI_{\mathrm{ref}}$,
	\item the data sources (IMF WEO) and update cadence (annual),
	\item the total maximum supply of KLD (10 billion),
	\item the permanent disablement of further minting,
	\item the irreversibility of burns,
	\item the requirement that all protocol actions are executed via multi-signature accounts.
\end{itemize}
KC7/index constitutional rules are not amendable; any alternative index requires a new token version.

	\noindent\textbf{Governable Economic Parameters (Within Fixed Bounds).}
	Certain monetary-policy parameters may be adjusted through governance, but only within predefined 
	ranges that preserve the structure of the mechanism. These include:
	\begin{itemize}
		\item issuance sensitivity $\alpha_I$,
		\item burn sensitivity $\beta_B$,
		\item escrow release multiplier $\alpha_E$,
		\item smoothing parameter $\gamma$,
		\item maximum burn fraction $b_{\max}$,
		\item staking or reward multipliers tied to protocol participation.
	\end{itemize}
	Each parameter has a published allowable range (e.g., $\alpha_I \in [0, 0.05]$), and governance 
	cannot exceed these bounds. Changes require a standard governance vote with a defined quorum and 
	majority threshold.
	
	\noindent\textbf{Protocol upgrades (new token version only).} The KC7 index definition, baseline construction, and the functional forms that map the debt index into issuance, burn, escrow, and staking rules are not amendable under KLD governance. Any proposal to alter these structural elements requires a new token version with a new genesis specification and independent audit. Governance may only adjust the governable parameters within fixed bounds as specified below.

    \noindent\textbf{Oracle Responsibilities.}
	Oracle operators have a strictly operational role. They retrieve the official data from the fixed
	sources, compute $X_t$ exactly as defined in the immutable specification, and publish the value
	on-ledger with a timestamp. They cannot modify the metric, the formula, the sources, or the update
	schedule. Governance may coordinate oracle membership and technical parameters, but not the
	economic definition of the index.
	
	\paragraph{Mechanics}
	\begin{itemize}[noitemsep]
		\item \textbf{Proposal submission}: accounts with minimum staked KLD can submit proposals.
		\item \textbf{Voting}: token-weighted voting, with snapshots to prevent manipulation.
		\item \textbf{Quorum}: default 5\% of circulating supply.
		\item \textbf{Voting period}: default 7 days.
		\item \textbf{Timelock}: default 48 hours before execution of approved changes.
	\end{itemize}
	\noindent\textbf{Voting Power Boundaries.}
	To ensure governance legitimacy and prevent capture, the following rules apply:
	\begin{itemize}
		\item \textbf{Treasury and Escrow Wallets.} Tokens held in protocol treasury or ecosystem 
		escrow accounts are excluded from governance voting. These allocations exist for programmatic 
		release and cannot be used to influence votes.
		
		\item \textbf{Unvested Team and Advisor Tokens.} Tokens subject to vesting schedules are 
		ineligible for voting until they are fully vested and circulating. This prevents insiders 
		from exercising disproportionate control during early stages.
		
		\item \textbf{Quorum Basis.} All quorum and majority thresholds are calculated relative to 
		the circulating and eligible supply only. Non‑circulating or ineligible tokens are excluded 
		from denominator calculations.
	\end{itemize}
	
	\section{Security and risk}
	
	\subsection{Technical security}
	
Treasury and issuer keys are managed by multisignature and kept in hardware security modules. Prior to deployment, control logic and escrow are audited. Suspicious flows on XRPL can be found with the aid of monitoring and anomaly detection.
	
	\subsection{Economic risk}
	
	The model mitigates economic risks by:
	\begin{itemize}[noitemsep]
		\item Imposing a hard cap \(S_{\max}\) on supply.
		\item Tying net issuance and burn tightly to debt via \(g_t\).
		\item Providing a clear framework for tightening or relaxing policy as macro conditions evolve.
	\end{itemize}
	
	\subsection{Oracle risk}
Governance may stop policy execution, replace or remove Oracle operators, postpone an update and keep the most recent confirmed value, or revert to a conservative default where no releases or parameter changes take place in the event of data outages, reporting errors, or obvious Oracle malfunction. Under no circumstances may governance change, override, or manually set the debt index's value.

	\subsection{Legal and compliance}
	
	KLD has a macro debt-indexed monetary policy and is intended to be a utility and governance token. The project may implement KYC/AML procedures for specific distributions and will seek legal opinions in pertinent jurisdictions. If necessary, the issuer can limit holdings to verified accounts in particular regions using XRPL's \texttt{RequireAuth} flag.
	
	\noindent\textbf{Holder Authorization (RequireAuth).} The MPT may be set up with the \texttt{RequireAuth} flag, which limits transfers to an allow-listed set of holders, if necessary for controlled distribution phases. Trustlines, which are unique to legacy issued currencies on XRPL, are not involved in this mechanism, which operates at the token-holder level.
	
	\section{Roadmap}
The project's roadmap is divided into the following four major phases:
	\paragraph{Phase 0: Model finalization and audits}
Completing the baseline policy, parameter ranges, and construction of the debt index. Examine the XRPL implementation design, Oracle architecture, and mathematical model.
	\paragraph{Phase 1: Issuance and escrow deployment}
Creating escrows for the escrow allocation and issuing $S_{\max}$ KLD on XRPL. Deploying oracles and the on-ledger publication + verification framework that links the published \(g_t\) value to multisignature policy execution (releases, burns, and escrow updates).
	\paragraph{Phase 2: Governance and integrations}
Introducing staking, governance, and fundamental utility integrations. We implement the debt-indexed policy in production and start routinely updating $g_t$ based on external data.
	\paragraph{Phase 3: Refinement and growth}
Based on feedback from the community and observed behavior, we adjust the parameters. We deepen liquidity, broaden integrations, and keep refining the model and Oracle infrastructure.\\
\noindent\textbf{Monetary policy refinement.}
The debt index definition, baseline ratio, maximum supply, and minting disablement are examples of immutable rules that are not altered by the protocol. Through standard governance procedures, governance can modify some governable parameters (such as $\alpha_I$, $\beta_B$, $\alpha_E$, $\gamma$, and $b_{\max}$) within predetermined bounds. Structural changes to the index definition, baseline construction, or policy functional forms require a new token version (not a governance amendment). Governance may only adjust listed parameters within fixed bounds.

	\section{Conclusion}
Kladia Liquidity Deflator (KLD) is a deflationary, debt-indexed token whose economic strategy is directly linked to an external indicator of the KC7 bloc debt. KLD provides a mathematically sound and practically feasible method of making token scarcity react to macro leverage by modeling debt using a normalized index $g_t$ and mapping this index into specific, bounded token-level actions. As the ecosystem and data landscape change, the model is intended to be transparent, auditable, and flexible through governance.
	
The KC7 bloc list is deliberately limited to seven high‑transparency economies: United States, Euro 
Area (EA20), Japan, United Kingdom, Canada, Australia, and South Korea. These blocs were chosen 
because \emph{(i)} they represent a large share of global GDP, \emph{(ii)} they publish debt and GDP statistics 
with consistent definitions and timely updates, and \emph{(iii)} their data is consolidated in the IMF World 
Economic Outlook (WEO) database, which provides a single canonical source. Restricting the index to 
this set ensures reproducibility and minimizes disputes over definitions, revisions, or missing data. 
The bloc list has the
option to introduce an “Extended” index (e.g., KE15) later without altering the KC7 baseline.

	\section*{Declaration of Interest}
	The authors have no conflicts of interest to declare. Both authors have seen and agree with the contents of the manuscript, and there is no financial interest to report.
	\section*{Data Availability Statement}
	Not applicable
	
	\bibliographystyle{unsrtnm}
	\bibliography{sn-bibliography}

	\vfill
	\noindent\textbf{Contact and further information}\\
	Please consult the project's official repositories and communication channels for technical information, implementation updates, and governance proposals.\\
	\begin{itemize}
		\item Website: \url{https://kladia.com}\\
	\end{itemize}



\end{document}